\documentclass[aps,prl,reprint,superscriptaddress,showpacs,amsmath,amssymb,longbibliography,lengthcheck]{revtex4-1}

\usepackage[dvipdfmx]{graphicx}
\usepackage{color}
\usepackage{mathrsfs}
\usepackage{txfonts}

\begin{document}

\title{Quantum Phase Transition in the Shape of Zr isotopes}

\author{
Tomoaki Togashi$^{1}$, Yusuke Tsunoda$^1$, Takaharu Otsuka$^{1,2,3,4}$ and Noritaka Shimizu
}

\affiliation{
  Center for Nuclear Study, University of Tokyo, Hongo, Bunkyo-ku, Tokyo 113-0033, Japan \\
  $^2$Department of Physics, University of Tokyo, Hongo, Bunkyo-ku, Tokyo 113-0033, Japan \\
  $^3$National Superconducting Cyclotron Laboratory,
  Michigan State University, East Lansing, Michigan 48824, USA\\
  $^4$Instituut voor Kern- en Stralingsfysica, KU Leuven, B-3001 Leuven, Belgium
}

\begin{abstract}
  The rapid shape change in Zr isotopes near neutron number $N$=60 is identified 
  to be caused by type II shell evolution associated with massive proton
  excitations to its $0g_{9/2}$ orbit, and is shown to be a quantum phase transition.    
  Monte Carlo shell-model calculations are carried out 
  for Zr isotopes of $N$=50-70 with many configurations spanned by eight proton orbits and eight neutron orbits. 
  Energy levels and B(E2) values are obtained within a single framework in a good agreement with experiments, 
  depicting various shapes in going from $N$=50 to 70.  
  Novel coexistence of prolate and triaxial shapes is suggested.
\end{abstract}


\maketitle

The shape of the atomic nucleus has been one of the primary subjects of nuclear structure physics 
\cite{bohr_mottelson}, and continues to provide intriguing and challenging questions in going to exotic nuclei.   
One such question is the transition from spherical to deformed shapes 
as a function of the neutron (proton) number $N$ ($Z$), referred to as {\it shape transition}.   
The shape transition is visible in the systematics of the excitation energies of low-lying states, 
for instance, the first 2$^+$ levels of even-even nuclei: it turns out to be high (low) for spherical (deformed) 
shapes \cite{bohr_mottelson,ring_schuck,casten}.  
A shell model (SM) calculation is suited, in principle, for its description, because of the high capability of 
calculating those energies precisely.   On the other hand, since the nuclear shape is a consequence of 
the collective motion of many nucleons, the actual application of the SM encountered 
some limits in the size of the calculation.

\begin{figure}[tbhp]
  \includegraphics[scale=0.28]{Fig1a.eps}\\
  \includegraphics[scale=0.28]{Fig1b.eps}\\
  \hspace{-1.5mm}
  \includegraphics[scale=0.28]{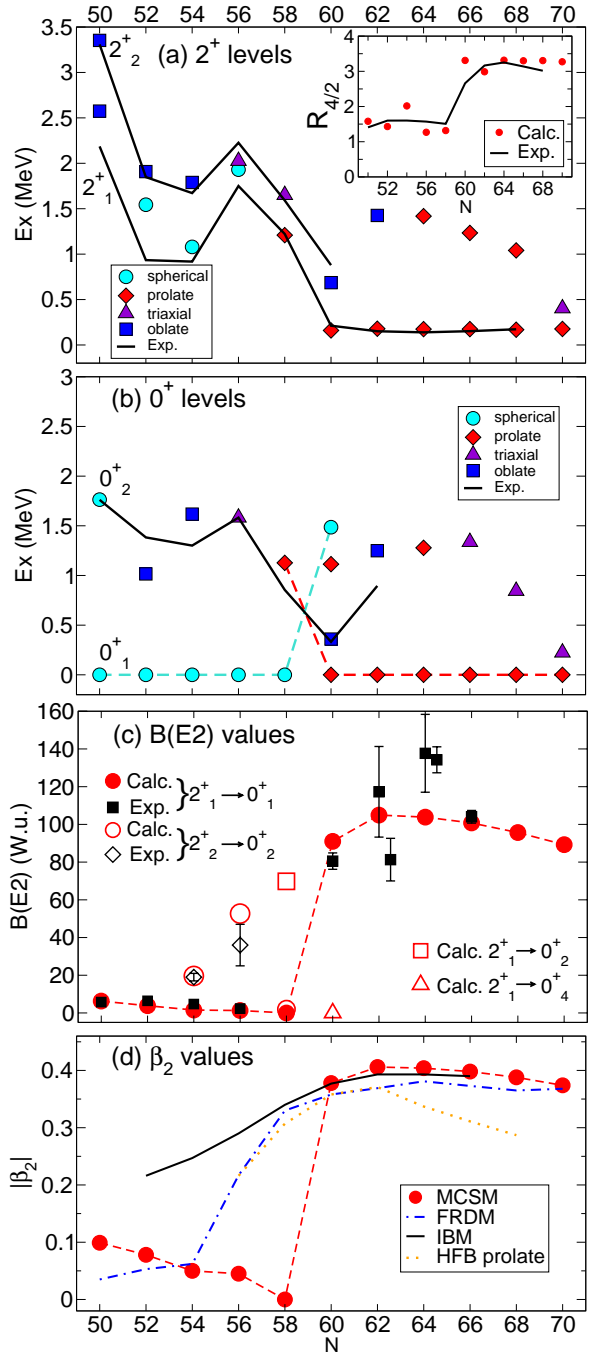}\\
  \hspace{-1.5mm}
  \includegraphics[scale=0.28]{Fig1d.eps}
  \caption{(Color online) 
    (a) $2^+_{1,2}$ levels, (b) $0^+$ levels of Zr isotopes as a function of $N$.
    Symbols are present theoretical results with the shape classification as shown in the 
    legends (see the text for details).  Solid lines denote experimental data   
    \cite{nudat2,ex100Zr1,ex100Zr2,ex100Zr3,ex102Zr,ex100102104Zr,ex104Zr,BE2_100104,
    ex104106Zr,ex106108Zr,ex108Zr}.
    Dashed lines connect relevant results to guide the eye.  
    The ratio between the $4^+_1$ and $2^+_1$ levels is shown in the insert of (a)
    in comparison to experiment.  The lowest four $0^+$ levels are shown for $^{100}$Zr.
    (c)  $B(E2; 2^+\rightarrow0^+)$ values as a function of $N$.  Experimental data are 
    from \cite{BE2,BE2_96,BE2_100104,BE2_102,BE2_104106,BE2_94Zr_2+2,kremer}.
    (d) Deformation parameter $\beta_2$.  The values by other methods are shown, too.
    }
  \label{sys}
\end{figure}

In this Letter, we present results of large-scale Monte Carlo Shell Model (MCSM) calculations 
\cite{mcsm_review1}  
on even-even Zr isotopes with a focus on the shape transition 
from $N =$ 50 to $N =$ 70, {\it e.g.} \cite{federman79}. 
Figure~\ref{sys}(a) shows that the observed 2$^+_1$ level moves up and down within the 1-2 MeV region 
for $N$=50-58, whereas it is quite low ($\sim$0.2 MeV) for $N\ge$ 60 
\cite{nudat2,ex100Zr1,ex100Zr2,ex100Zr3,ex102Zr,ex100102104Zr,ex104Zr,BE2_100104,ex104106Zr,
ex106108Zr,ex108Zr}.
Namely, a sharp drop by a factor of $\sim$6
occurs at $N$=60, which is consistent with the corresponding B(E2) values shown 
in Fig.~\ref{sys}(c).  These features have attracted much attention, also because no theoretical 
approach seems to have reproduced those rapid changes covering both sides.  More importantly, 
an abrupt change seems to occur in the structure of the ground state as a function of $N$, which can be 
viewed as an example of the quantum phase transition (QPT) satisfying its general definition 
to be discussed \cite{qpt1,qpt2}.
This is quite remarkable, as the shape transition is in general rather gradual.  
In addition, there is much interest in those Zr isotopes from the viewpoint of the shape 
coexistence \cite{coexist_review1}.  

\setlength{\arrayrulewidth}{0.45pt}
\begin{table}[bthp]
\caption{Model space for the shell model calculation.}
\label{mspace}
\begin{center}
\begin{tabular}{|c|c|c|}
\hline
\,proton\,orbit\,\,&\,\,magic\,number\,\,&\,\,neutron\,orbit\,\, \\
\hline
-& &$1f_{7/2},2p_{3/2}$\\
\hline
 &82& \\
\hline
-& &$0h_{11/2}$\\
$0g_{7/2},1d_{5/2,3/2},2s_{1/2}$& &$0g_{7/2},1d_{5/2,3/2},2s_{1/2}$\\
\hline
 &50& \\
\hline
$0g_{9/2}$& &$0g_{9/2}$\\
$0f_{5/2},1p_{3/2,1/2}$& &-\\
\hline
\end{tabular}
\end{center}
\end{table}

The advanced version of MCSM \cite{mcsm_ptep, mcsm_reorder} 
can cover all Zr isotopes in this range of $N$ with a fixed Hamiltonian, when taking a large model space, 
as shown in Table~\ref{mspace}.   
The MCSM, thus, resolves the difficulties of conventional SM calculation, where the largest dimension reaches 
3.7$\times$10$^{23}$, much beyond its current limit.   
Note that no truncation on the occupation numbers of these orbits is made in the MCSM.  
The structure of Zr isotopes has been studied by many different models and theories.
For instance, a recent large-scale conventional SM calculation showed a rather accurate reproduction of experimental data up to $N$=58, whereas it was not extended beyond $N$=60 \cite{sieja09}.    
The 2$^+_1$ levels have been calculated in a wider range in Interacting Boson Model (IBM) calculations, 
although the afore-mentioned rapid change is absent \cite{ibm_1,ibm_2}.   
Some other works were restricted to deformed states \cite{federman79,xu02,psm}, 
or indicated gradual shape-changes 
\cite{vampir,hfb,hfb+GCM,xiang,RMF,moller95,skalski,smmc}.

It is, thus, very timely and needed to apply the MCSM to Zr isotopes, particularly heavy exotic ones.
The Hamiltonian of the present work is constructed from existing ones, so as to reduce 
ambiguities.  
The JUN45 Hamiltonian is used for the orbits, $0g_{9/2}$ and below it \cite{jun45}.
The SNBG3 Hamiltonian \cite{snbg3} is used for the $T$=1 interaction for $0g_{7/2}$, $1d_{5/2,3/2}$, 
$2s_{1/2}$ and $0h_{11/2}$.  Note that the JUN45 and SNBG3 interactions were obtained
by adding empirical fits to microscopically derived effective interactions \cite{jun45,snbg3}.  
The $V_{{\rm MU}}$ interaction \cite{vmu} is taken for the rest of the effective interaction.  
The $V_{{\rm MU}}$ interaction consists of the central part given by a Gaussian 
function in addition to the $\pi$- and $\rho$-meson exchange 
tensor force \cite{vmu}.  The parameters of the central part were fixed from monopole 
components of known SM interactions \cite{vmu}.
The $T$=0 part of the $V_{{\rm MU}}$ interaction is kept unchanged throughout this work. 
The $T$=1 central part is 
reduced by a factor of 0.75 except for $1f_{7/2}$ and $2p_{3/2}$ orbits.
On top of this, $T$=1 two-body matrix elements 
for $0g_{9/2}$ and above it, including those given by the SNBG3 interaction, are 
fine tuned by using the standard method \cite{honma02,brown06}.
The observed levels of the 2$^+_1$ and 4$^+_1$ states of 
$^{90-96}$Zr and the 0$^+_2$ state of $^{94-100}$Zr are then used.  Since the number of available
data is so small, this cannot be a fit but a minor improvement.  
The single-particle energies are determined so as to be consistent with the prediction of the 
JUN45 Hamiltonian, the observed levels of $^{91}$Zr with spectroscopic factors, {\it etc}.  
The present SM Hamiltonian is, thus, fixed, and no change is made throughout 
all the calculations below.   It is an initial version, and can be refined for better details.   

Figure \ref{sys}(a) shows excitation energies of the 2$^+_{1,2}$ states of the Zr isotopes, indicating   
that the present MCSM results reproduce quite well the observed trends. 
The shape of each calculated state is assigned as spherical, prolate, triaxial or oblate by the method of 
\cite{mcsm_ni68}, as will be discussed later.
The calculated 2$^+_1$ state is spherical for $N$=52-56, while it becomes prolate deformed for $N\ge$58. 
Its excitation energy drops down at $N$=60 by a factor of $\sim$6, and stays almost constant, in agreement 
with experiment.  
The ratio between the $4^+_1$ and $2^+_1$ levels, denoted $R_{4/2}$, is depicted in the insert of 
Fig.~\ref{sys}(a) in comparison to experiment.  
The sudden increase at $N$=60 is seen in both experiment and calculation,
approaching the rotational limit, 10/3, indicative of a rather rigid deformation.
The $R_{4/2} < 2$ for $N\le$58 suggests a seniority-type structure which stems from 
the $Z$=40 semi-magicity.  
 
\begin{figure*}[tb]
\includegraphics[scale=0.50]{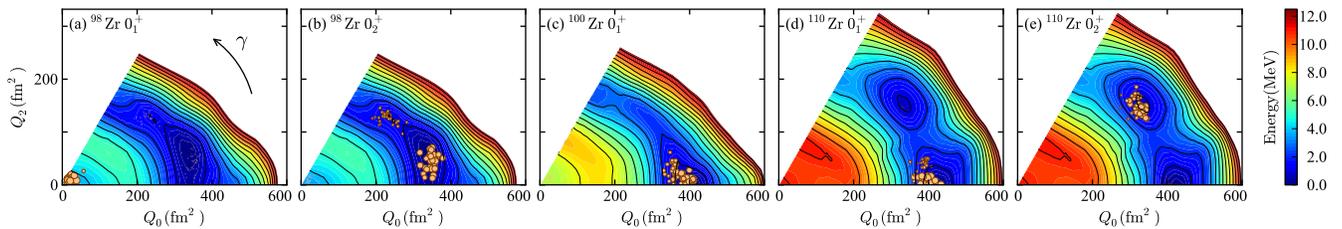}
\caption{ (Color online) 
{\it T-plots} for $0^+_{1,2}$ states of $^{98,100,110}$Zr isotopes. }
\label{tplot}
\end{figure*}

Figure \ref{sys}(b) shows the properties of $0^+_{1,2}$ states.  Their shapes are  
assigned in the same way as the 2$^+$ states.  The ground state remains spherical up to
$N$=58, and becomes prolate at $N$=60.   A spherical state appears as the 
$0^+_4$ state at $N$=60 instead, as shown in Fig.~\ref{sys}(b). 
We here sketch how the shape assignment is made for the MCSM eigenstate.  
The MCSM eigenstate is a superposition of MCSM basis vectors projected onto the 
angular momentum and parity.   Each basis vector is a Slater determinant, {\it i.e.}, 
a direct product of superpositions over original single-particle states.  
The optimum amplitudes in such superpositions 
are searched based on quantum Monte-Carlo and variational methods \cite{mcsm_review1,mcsm_ptep}. 
For each MCSM basis vector so fixed, we can compute and diagonalize its quadrupole matrix.  
This gives us the three axes of the ellipsoid 
with quadrupole momenta $Q_0$ and $Q_2$ in the usual way \cite{ring_schuck}. 
One can then plot this MCSM basis vector as a circle on the Potential Energy Surface (PES) 
, as shown in Fig.~\ref{tplot}.  
The overlap probability of this MCSM basis vector with the eigenstate is indicated
by the area of the circle.  Thus, one can pin down each MCSM basis vector on the PES 
according to its $Q_0$ and $Q_2$ with its importance by the area of the circle.  
Note that the PES in Fig.~\ref{tplot} is obtained by constrained HF calculation for the 
same SM Hamiltonian, and is used for the sake of an intuitive understanding of MCSM results.     
This method, called a {\it T-plot} \cite{mcsm_ni68}, enables us to analyze SM eigenstates
from the viewpoint of intrinsic shape. 
Figure \ref{tplot}(a) shows that the MCSM basis vectors of the $0^+_1$ state of $^{98}$Zr 
are concentrated in a tiny region of the spherical shape, while its $0^+_2$ state is composed 
of basis vectors of prolate shape with $Q_0\sim$350 fm$^2$ (see Fig.~\ref{tplot}(b)).  
A similar prolate shape dominates the $0^+_{1}$ state of $^{100}$Zr with slightly larger $Q_0$, 
as shown in Fig.\ref{tplot}(c).  We point out the abrupt change of the ground-state property
from Fig.~\ref{tplot}(a) to (c), and will come back to this point later.
The {\it T-plot} shows stable prolate shape for the $0^+_{1}$ state from $^{100}$Zr to 
$^{110}$Zr (see Fig.~\ref{tplot}(d)).   

Figure \ref{sys}(c) displays $B(E2; 2^+_1\rightarrow0^+_1)$ values, 
with small values up to $N$=58 and a sharp increase at $N$=60, consistent with experiment
\cite{BE2,BE2_96,BE2_100104,BE2_102,BE2_104106}.
The effective charges, $(e_p, e_n) = (1.3e, 0.6e)$, are used.  
Because the $B(E2; 2^+_1\rightarrow0^+_1)$ value is a sensitive probe of the 
quadrupole deformation, 
the salient agreement here implies that the present MCSM calculation produces quite well 
the shape evolution as $N$ changes.  
In addition, theoretical and experimental $B(E2; 2^+_2\rightarrow0^+_2)$ values are shown for $N$=54 
\cite{BE2_94Zr_2+2} and 56.
The value for $N$=56 has been measured by experiment, discussed in the subsequent paper \cite{kremer}, 
as an evidence of the shape coexistence in $^{96}$Zr.
The overall agreement between theory and experiment appears to be remarkable.
It is clear that the $2^+_2\rightarrow0^+_2$ transitions at $N$=54 and 56 are linked to the 
$2^+_1\rightarrow0^+_1$ transitions in heavier isotopes, via $2^+_1\rightarrow0^+_2$ transition at $N$=58.

Figure \ref{sys}(d) shows the deformation parameter $\beta_2$ \cite{bohr_mottelson}.  
The results of IBM \cite{ibm_2}, HFB \cite{hfb} and FRDM \cite{moller95} calculations are included, exhibiting much more gradual changes. The MCSM values are obtained from $B(E2; 2^+_1\rightarrow0^+_1)$.

The systematic trends indicated by the $2^+_1$ level, the ratio $R_{4/2}$, the  
$B(E2; 2^+_1\rightarrow0^+_1)$ value (or $\beta_2$), and the {\it T-plot} analysis are all consistent among themselves and in agreement with relevant experiments.  We can, thus, identify the change 
between $N$=58 and 60 as a QPT, 
where in general an abrupt change should occur in the quantum structure of the ground state for a 
certain parameter \cite{qpt1,qpt2}.  The parameter here is nothing but the neutron number $N$, and the transition occurs from a ``spherical phase'' to a ``deformed phase''.  
Figure \ref{sys}(b) demonstrates that the 0$^+_1$ state is spherical up to $N$=58, but the  
spherical 0$^+$ state is pushed up to the 0$^+_4$ state at $N$=60, where the prolate-deformed 0$^+$ state
comes down to the ground state from the 0$^+_2$ state at $N$=58.  This sharp crossing causes the present QPT.  
The discontinuities of various quantities, one of which can be assigned the order parameter, 
at the crossing point imply the first-order phase transition.   
The shape transition has been noticed in many chains
of isotopes and isotones, but appears to be rather gradual in most cases, for instance,
from $^{148}$Sm to $^{154}$Sm.  The abrupt change in the Zr isotopes is exceptional. 

We comment on the relation between the QPT and the modifications of the interaction
mentioned above.  Without them, the 2$^+_1$ level is still $\sim$0.2 MeV at $N$=60 close to Fig.~\ref{sys}(a), 
while at $N$=58 it is higher than the value in Fig.~\ref{sys}(a).  
Thus, the present QPT occurs rather insensitively to the modifications, whereas experimental data
can be better reproduced by them. 

\begin{figure}[tb]
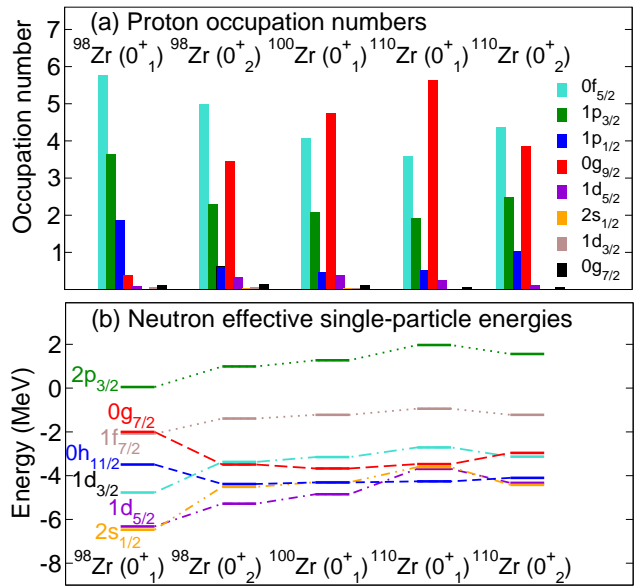

\includegraphics[scale=0.30]{Fig3a.eps}\\
\vspace{0.1cm}
\includegraphics[scale=0.30]{Fig3b.eps}
\caption{ (Color online)
(a) Occupation numbers of protons and (b) effective single-particle energies of neutrons
for selected Zr isotopes. Neutron $0g_{9/2}$ is around -12 MeV, and is not shown.}
  \label{occ_espe}
\end{figure}

We now discuss the origin of such abrupt changes.
Figure \ref{occ_espe}(a) displays the occupation numbers of proton orbits 
for the $0^+_{1,2}$ states of $^{98}$Zr, the $0^+_{1}$ state of $^{100}$Zr and 
the $0^+_{1,2}$ states of $^{110}$Zr.
From the spherical $0^+_{1}$ to prolate $0^+_{2}$ states of $^{98}$Zr,  
the occupation number of the proton $0g_{9/2}$ increases from 0.4 to 3.5, 
while those of the $pf$-shell orbits decrease.  The proton $0g_{9/2}$ 
orbit is more occupied in the prolate $0^+_{1}$ state of $^{100,110}$Zr.

Figure \ref{occ_espe}(b) shows effective single-particle energies (ESPE) of neutron orbits 
calculated with the occupation numbers of the 
SM eigenstates, shown in Fig.~\ref{occ_espe}(a) (see \cite{otsuka16,mcsm_ni68} for explanations).
At a glance, one notices that the ESPEs from $2s_{1/2}$ to $0g_{7/2}$ are distributed over a range 
of 4 MeV for the $0^+_{1}$ state of $^{98}$Zr, but are within 2 MeV for the prolate states, such as 
$0^+_{2}$ of $^{98}$Zr, $0^+_{1}$ of $^{100}$Zr and 
$0^+_{1}$ of $^{110}$Zr.
We notice also a massive (3.5-5.5) excitation of protons into $0g_{9/2}$ 
in these prolate states (see Fig.~\ref{occ_espe}(a)).   
These two phenomena are correlated, and are, indeed, predicted in the type II shell evolution scenario 
\cite{mcsm_ni68,otsuka16}, where particular particle-hole excitations can vary the shell structure significantly.
(See Ref.~\cite{otsuka16} for an overview of type I and II shell evolutions, and Ref.~\cite{kremer} for the discussion 
on $^{96}$Zr.)
To be more concrete, protons in the $0g_{9/2}$ orbital lower the ESPEs of neutron $0g_{7/2}$ and $0h_{11/2}$
orbitals more than other orbits.
For the $0g_{9/2}$-$0g_{7/2}$ coupling, the tensor and central forces work coherently \cite{tensor,vmu,nobel}, 
and substantial lowering ($\sim$2 MeV) occurs. In the $0g_{9/2}$-$0h_{11/2}$ case, the tensor and central forces 
work destructively but the net effect is still lowering, though weaker than the other case. 
Regarding the central force, the attraction between unique-parity 
orbits is stronger than the average due to similarities in radial wave functions, as also mentioned earlier 
by Federman and Pittel \cite{federman_pittel,otsuka16}.
The present deformation is primarily a result of the quadrupole component of the effective interaction, and 
is enhanced by coherent contributions of various configurations 
(Jahn-Teller effect  \cite{jahn_teller}).  
If single-particle energies are spread with sizable gaps in between, such coherence is disturbed and 
the deformation is suppressed.
In the present prolate states, by distributing protons and neutrons in a favorable way partly by 
particle-hole excitations, 
ESPEs can be optimized for stronger deformation as much as possible, thanks to the monopole 
properties of the central and tensor forces \cite{tensor,vmu,nobel}.  
This is the idea of type II shell evolution \cite{otsuka16,mcsm_ni68}, and one finds that it occurs here.

Such reorganization of the shell structure involves substantial re-configuration of 
protons and neutrons (or type II shell evolution), 
leading to more different configurations 
between the normal states and the states with this deformation-optimized shell structure.
This property results in a suppressed mixing of two such states even around their crossing point. 
The abrupt change, thus, appears with almost no mixing, leading to a QPT. 
In order to have such a situation, a unique-parity orbit, like $0g_{9/2}$, should sit 
just above a closed shell.  This can be fulfilled in the $_{38}$Sr isotopes to a certain extent with 
similar but less distinct systematic changes.   In other elements, however, there is no such case
known so far, making the Zr (and Sr) isotopes quite unique at this time.  In fact, 
other cases with somewhat weaker effects of type II shell evolution turn out to be shape coexistence 
in various forms.  For instance, in $^{68}$Ni case, the proton $pf$ shell plays a similar role to the
present neutron orbits, but has somewhat weaker collectivity \cite{mcsm_ni68,otsuka16}. 

\begin{figure}[tb]
\includegraphics[scale=0.30]{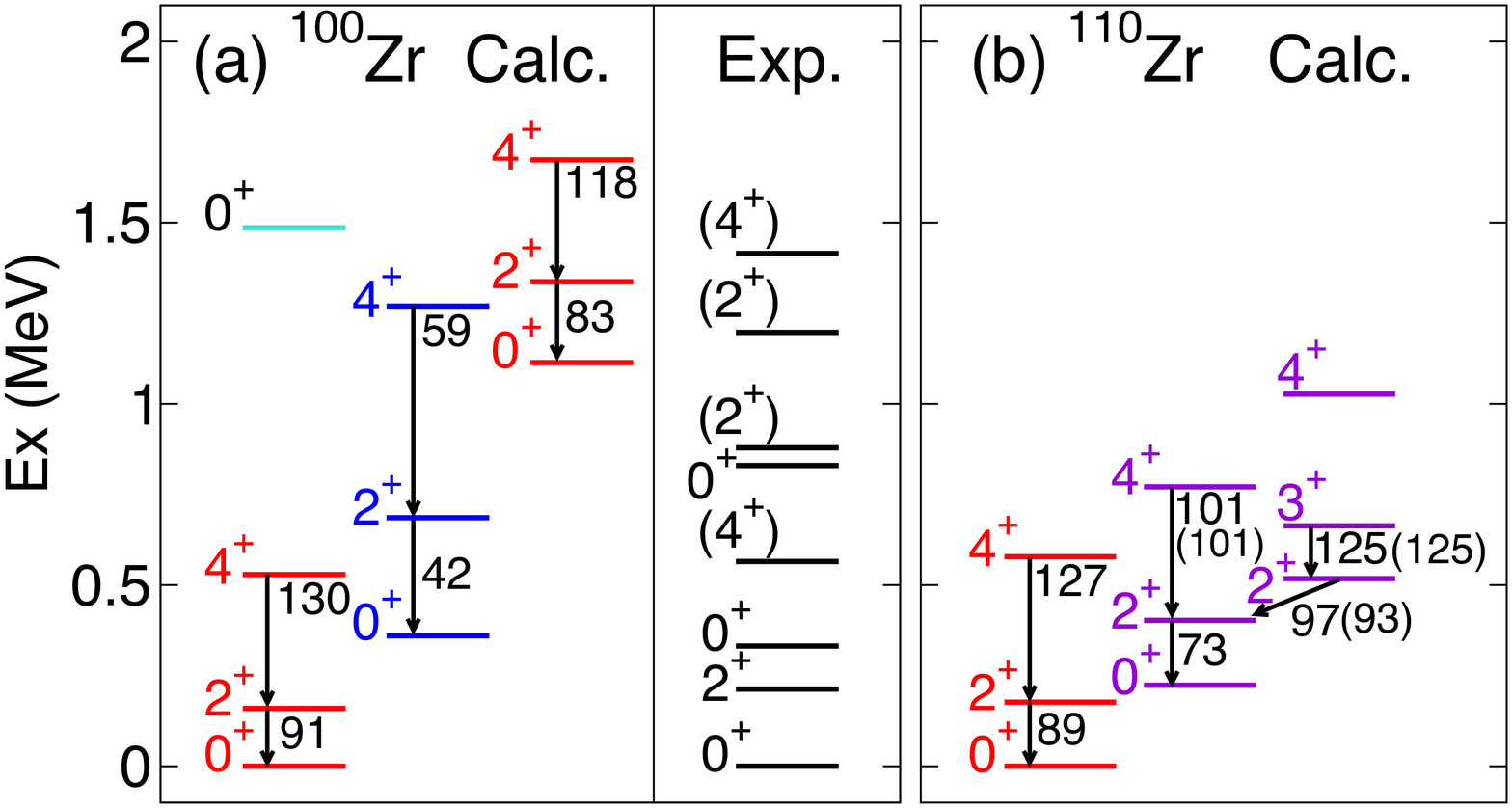}
\caption{ Levels of (a) $^{100}$Zr and (b) $^{110}$Zr.  Prolate, oblate and triaxial
bands are shown in red, blue and purple, respectively.
The $0^+_4$ state in light blue in (a) is spherical.
Some large B(E2) values are shown in W.u., with rigid-triaxial-rotor values in parentheses.
 }
\label{level}
\end{figure}

Figure \ref{level} indicates that the prolate ground bands are similar 
between $^{100}$Zr and $^{110}$Zr, but an intriguing difference appears in side bands.  
Figure \ref{level}(a) depicts the coexistence of the prolate and oblate bands  
with reasonable agreement to experiment.
The excited band of $^{110}$Zr corresponds to a triaxial shape with a profound local minimum 
at $\gamma \sim 30 ^\circ$ in Fig.~\ref{tplot}(e).  It co-exists 
with the prolate band in such a close energy, because their
ESPEs are so different (see Fig.~\ref{occ_espe}(b))
due to different proton occupations shown in Fig.~\ref{occ_espe}(a). 
Note that neutron ESPEs for the $0^+_2$ have two substructures with a gap between $0h_{11/2}$ and $1d_{3/2}$.
The B(E2) values in this triaxial band are almost identical to those given by 
the rigid-triaxial rotor model of Davydov and Filippov
with $\gamma$=28$^\circ$ \cite{DFtriaxial_1,DFtriaxial_2}.  
Their prediction normalized by the $B(E2; 2^+_2\rightarrow0^+_2)$ value is included in Fig.~\ref{level}(b).  
Type II shell evolution thus produces another interesting case.
The transition from Figs.~\ref{tplot}(c) to (d,e) suggests a possible second-order phase transition 
at larger $N$ values, as a future issue.

In summary, a quantum phase transition of the nuclear shape has been 
shown to occur in the Zr isotopes.  The abrupt change appears with a fixed Hamiltonian
through type II shell evolution.  The re-organization of the shell structure due to type II shell 
evolution provides us with a new way to look into nuclear structure, and is expected to occur in other nuclei.  
The lowest states of these Zr isotopes 
provide a variety of shapes and their coexistence 
(see Ref.~\cite{kremer} for $^{96}$Zr), including a novel situation of
prolate-triaxial coexistence.  Further investigations, for instance on octupole shapes, are of much
interest, {\it e.g.} \cite{mach90,kremer}. 

\section*{Acknowledgement}
\vspace{-2mm}
We thank Prof. S.~Miyashita for valuable comments on the QPT.
We are grateful to Prof. B.R. Barrett and Prof. P. Van Duppen for useful remarks, 
and to Prof. N. Pietralla and Dr. C. Kremer for various discussions, 
including those on their experimental data prior to publication.
This work was supported in part by Grants-in-Aid
for Scientific Research (23244049). 
It was supported in part by HPCI Strategic Program (hp150224),
in part by MEXT and JICFuS
and a priority issue (Elucidation of the fundamental laws and
evolution of the universe)
to be tackled by using Post ``K'' Computer (hp160211),
and by CNS-RIKEN joint project for large-scale nuclear structure
calculations.

\end{document}